\documentclass[twocolumn,floatfix,nofootinbib,prd,preprintnumbers,showpacs,showkeys,superscriptaddress,longbibliography]{revtex4-1}

\usepackage{amsmath,amssymb,bm,graphicx}
\usepackage{xcolor}
\usepackage[colorlinks=true,linkcolor=blue,citecolor=magenta,urlcolor=blue]{hyperref}
\usepackage{orcidlink} 
\usepackage{comment}

\newcommand{\dd}{\mathrm{d}}
\newcommand{\ee}{\mathrm{e}}
\newcommand{\ii}{\mathrm{i}}
\newcommand{\eps}{\epsilon}

\newcommand{\W}{W}
\newcommand{\q}{q} 
\newcommand{\beq}{\begin{equation}} 
\newcommand{\eeq}{\end{equation}} 
\newcommand{\St}{S}
\renewcommand{\boxed}[1]{#1}
\begin{document}

\title{Planckian Gravitons from an Imaginary-Time Clock}

\author{Michael R.R. Good\,\orcidlink{0000-0002-0460-1941}}
\email{muon@asu.edu}
\affiliation{Physics Department \& Energetic Cosmos Laboratory, Nazarbayev University,\\
Astana 010000, Qazaqstan.}
\affiliation{Leung Center for Cosmology \& Particle Astrophysics,
National Taiwan University,\\ Taipei 10617, Taiwan.}
\affiliation{Beyond Center for Fundamental Concepts in Science, Arizona State University,\\
Tempe AZ 85287, USA.}

\author{Eric V.\ Linder\,\orcidlink{0000-0001-5536-9241}  }
\email{evlinder@lbl.gov} 
\affiliation{Berkeley Center for Cosmological Physics \& Berkeley Lab, University of California,\\ Berkeley CA 94720, USA.}

\begin{abstract} 
We present a simple derivation of the exact Planck spectrum of the quadrupole radiation from point masses moving apart nonrelativistically, essentially an analog for gravitational radiation. The standard Einstein quadrupole radiation formula gives emitted power proportional to the square of the third derivative of $x(t)^2$. In our moving-mass picture, imaginary-time periodicity appears as a product-log trajectory of a quadrupole source. In the frequency domain, the power becomes proportional to the Planck distribution, $\omega^3/(e^{2\pi c\omega/\kappa}-1)$.  The resulting Planckian graviton energy spectrum has finite total energy and finite graviton number.  The emitted spectrum is purely kinematic in origin: no equilibrium, horizon, or stochastic source is assumed.  

\end{abstract}

\keywords{gravitons, gravitational radiation, quadrupole formula, Planck spectrum, acceleration thermality}
\pacs{04.30.Db, 04.60.-m, 04.62.+v}
\date{\today}

\maketitle

\section{Introduction}
\label{sec:intro}

Planckian spectral factors are usually associated with equilibrium thermodynamics \cite{Planck1901}, but they can also arise from the kinematic and analytic structure of an accelerated classical source.  This note illustrates out-of-equilibrium kinematic thermal behavior in gravitational radiation. The effect is somewhat reminiscent of acceleration thermality of the Davies-Unruh effect \cite{Davies:1974th,Fulling:1972md, Davies:1974th,unruh76} which connects an accelerated observer to Planck radiation, or how the Hawking effect \cite{Hawking:1974sw,Hawking:1974rvNATURE} results in a Planck spectrum from a black hole, which can emerge kinematically from Bogoliubov transformations between inequivalent vacua in the presence of a spacetime horizon. 

Here, we consider a rectilinear source history (non-relativistic) and use the leading gravitational-wave quadrupole formula.  The goal is to isolate the solvable kinematic mechanism by which emitted gravitons acquire the Planck energy spectrum.

The Davies-Fulling effect \cite{DeWitt:1975ys,Davies:1976hi,Davies:1977yv} for scalar particle production from a moving mirror (accelerated boundary; see many recent works, e.g., \cite{
PisinChen2017, Reyes:2021npy, Agullo:2025opy, Hsiang:2024xlh, Xie:2023wvu, akal2022zoo, ievlev2024moving, Biswas:2024mlq, Kumar:2023kse, Dodonov:2025rxz, Lin:2021bpe, Gorban:2024vss, Chen:2020sir, Chen:2015bcg, AnaBHEL:2022sri, Navick:2024wbd, Hsiung:2025mya, Chen:2025xkv,Tomonaga:2024eoc}) depends on the time-dependent proper acceleration, and electrons emit the usual Larmor power \cite{Larmor1897}, which likewise depends on the proper acceleration \cite{Zhakenuly:2021pfm}, radiating Planckian photons in a dipole pattern (e.g., $\beta$-decay \cite{Ievlev:2023inj,Lynch:2022rqx,Good:2022xin,Good:2022eub,Good:2025qta}). However, gravitational radiation does not respond to dipole acceleration in the same way electromagnetic radiation does.  In the non-relativistic limit, the leading gravitational source is the mass quadrupole.  For rectilinear motion, this source quantity depends on a single scalar,
\begin{equation}
        \q(t)\equiv x(t)^2,
        \label{eq:qdef_intro}
\end{equation}
and the power is proportional to $[\dddot \q(t)]^2$.  Thus Planckian graviton emission can be generated by choosing the quadrupole history, rather than the acceleration (as in the photon case), to have the appropriate Fourier structure. 

The paper is organized as follows.
Section~\ref{sec:rectilinear} reduces the quadrupole formula for rectilinear nonrelativistic motion to a scalar source, $\dd^3 x(t)^2/\dd t^3$, and connects its Fourier transform to the energy spectrum.
Section~\ref{sec:trajectory} motivates the logarithmic imaginary-time clock and constructs the regularized product-log trajectory, source, and kinematic scale $\kappa$.
Section~\ref{sec:spectrum} derives the exact Planckian graviton spectrum, the total emitted energy, the temperature scale, and the graviton count. Section~\ref{sec:deformations} studies a one-parameter deformation of the logarithmic clock and identifies the product-log case as the particular exact Planckian member.
Section~\ref{sec:discussion} interprets the result as a kinematic, non-equilibrium frequency-space effect, compares it to electromagnetic radiation, connects it to entanglement entropy, and computes the 
angular radiation pattern.
Section~\ref{sec:concl} concludes. We use SI units throughout.

\section{Rectilinear quadrupole power}
\label{sec:rectilinear}

The leading nonrelativistic gravitational-wave power is the Einstein quadrupole formula \cite{Einstein:1918btx},
\begin{equation}
        P(t)=\frac{G}{5c^5}\sum_{i,j}\left(\dddot I_{ij}\right)^2,
        \label{eq:quadpower}
\end{equation}
where
\begin{equation}
        I_{ij}=m\left(x_i x_j-\frac{1}{3}r^2\delta_{ij}\right),
        \label{eq:quadmoment}
\end{equation}
is the trace-free mass quadrupole moment.  For rectilinear motion along the $x$-axis,
\begin{equation}
        \mathbf r(t)=(x(t),0,0),
        \label{eq:worldline_nr}
\end{equation}
so that
\begin{equation}
        I_{xx}=\frac{2}{3}m x^2,\qquad
        I_{yy}=I_{zz}=-\frac{1}{3}m x^2,
        \label{eq:rectI}
\end{equation}
with all off-diagonal components zero.  Hence
\begin{align}
        \sum_{i,j}\left(\dddot I_{ij}\right)^2
        &=
        \left(\frac{2m}{3}\right)^2
        \left[\frac{\dd^3}{\dd t^3}x^2\right]^2
        +2\left(\frac{m}{3}\right)^2
        \left[\frac{\dd^3}{\dd t^3}x^2\right]^2     \nonumber\\
        &=\frac{2m^2}{3}
        \left[\frac{\dd^3}{\dd t^3}x^2\right]^2 .
        \label{eq:sumreduction}
\end{align}
Substitution into Eq.~\eqref{eq:quadpower} gives
\begin{equation}
        \boxed{
        P(t)=\frac{2Gm^2}{15c^5}
        \left[\frac{\dd^3}{\dd t^3}x(t)^2\right]^2
        } .
        \label{eq:rectpower}
\end{equation}
Equivalently,
\begin{equation}
        \frac{\dd^3}{\dd t^3}\,x^2
        =6\dot x\ddot x+2x\dddot x.
        \label{eq:x2third}
\end{equation}
Thus, unlike for 
electromagnetic dipole radiation, the
rectilinear gravitational power is not proportional to \(\ddot x^2\).
It is governed instead by the third derivative of the quadrupole coordinate \(x^2\).

Let us define the source
\begin{equation}
    \St(t)\equiv \frac{\dd^3}{\dd t^3}x(t)^2,
\end{equation}
         and use the symmetric Fourier convention
\begin{align}
                \widehat{\St}(\omega)&=
        \frac{1}{\sqrt{2\pi}}
        \int_{-\infty}^{\infty}\St(t)e^{-i\omega t}\,\dd t,
        \nonumber\\
        \St(t)&=
        \frac{1}{\sqrt{2\pi}}
        \int_{-\infty}^{\infty}\widehat{\St}(\omega)e^{i\omega t}\,\dd\omega .
        \label{eq:source_def}
\end{align}
Since \(\St(t)\) is real, the associated positive-frequency energy spectrum is
\begin{equation}
        \boxed{
        \frac{\dd E}{\dd\omega}
        =
        \frac{4Gm^2}{15c^5}
        \left|\widehat{\St}(\omega)\right|^2
        } .
        \label{eq:spectrumgeneral}
\end{equation}
With the symmetric Fourier convention in Eq.~\eqref{eq:source_def}, Parseval's theorem gives
\begin{equation}
        \int_{0}^{\infty}\frac{\dd E}{\dd\omega}\,\dd\omega
        =
        \int_{-\infty}^{\infty}P(t)\,\dd t .
        \label{eq:parsevalcheck}
\end{equation}

\section{A Planckian quadrupole trajectory}
\label{sec:trajectory}
To motivate the quadrupole source history (or $x(t)$ trajectory), we first demand the analytic structure associated with a Planck denominator (i.e.\ the Bose-Einstein phase space distribution), and then ask what quadrupole history realizes it.  Let
\begin{equation}
        u=\frac{\kappa t}{c},
\end{equation}
and let \(y \sim x^2 >0\) denote the dimensionless quadrupole coordinate, proportional
to \(x^2\). As in the Hawking \cite{Hawking:1974sw,Hawking:1974rvNATURE} black hole temperature case, the Planck denominator for Euclidean thermal periodicity is tied to complex time, with \(\kappa\) setting the corresponding inverse-temperature scale \cite{Hartle:1976tp}.  The winding number counts repeated circuits around the complex
branch point, each circuit contributing one such thermal period \cite{Gibbons:1976ue}. Imaginary-time periodicity with period \(2\pi c/\kappa\) suggests that the inverse map \(u=u(y)\) contains a logarithmic branch,
because writing
\[
y = r e^{i\theta},
\]
gives
\[
\ln y = \ln r + i\theta .
\]
Taking \(y\) once around the origin in the complex plane corresponds to
\[
\theta \to \theta + 2\pi ,
\]
and therefore
\[
\ln y \to \ln r + i(\theta+2\pi)
=
\ln y + 2\pi i .
\]
Thus, the logarithm acquires an imaginary-time shift after one circuit around
the branch point at \(y=0\), i.e.,\ $\ln y \rightarrow \ln y+2\pi i$. 

The most direct trial is therefore a pure logarithm,
\begin{equation}
        u=\ln y,
        \qquad \Longrightarrow \qquad
        y=\ee^u .
\end{equation}
This has the desired imaginary-time shift, but it is not an acceptable
quadrupole history because \(y'''(u)=e^u\), so the source grows at late times and its Fourier transform is not defined as an ordinary convergent integral. 
The logarithm gives the branch structure, but by itself does not give late-time decay (i.e.,\ at large $y$).  

The simplest repair is to keep the same logarithmic branch while adding
one linear term to the inverse map:
\begin{equation}
        y+\ln y=u .
        \label{eq:inverse1}
\end{equation}
The logarithmic continuation still shifts \(u\) by \(2\pi i\), while the source now decays in both the past:
\begin{equation}
        y'''(u)\sim \ee^u \quad (u\to-\infty),
        \end{equation}
        and the future \begin{equation}
        y'''(u)\sim -\frac{2}{u^3} \quad (u\to+\infty).
\end{equation}
Thus, the Fourier integral is well-defined upon adding a linear term.  Equivalently, the Fourier phase
takes the gamma-transform form, 
with $\Omega\equiv \omega c/\kappa$, 
\begin{equation}
        \ee^{-\ii\Omega u}
        =
        \ee^{-\ii\Omega y}y^{-\ii\Omega},
\end{equation}
so the logarithmic monodromy is retained while the source has the decay
needed for a finite spectrum.

Solving Eq.~\eqref{eq:inverse1} gives
\begin{equation}
        y(t)=\W(\ee^u),
        \label{eq:ydef}
\end{equation}
where \(\W\) is the principal Lambert-\(W\) function, also called the
product log.  Restoring dimensions, 
we are led to the rectilinear trajectory\footnote{
The trajectory should be understood as a prescribed nonrelativistic
quadrupole history.  A center-of-mass-fixed realization is given by two
masses with positions \(\mathbf r_\pm(t)=(\pm x(t),0,0)\), for which the
quadrupole has the same form with \(m\) interpreted as the total mass.
The one-body framing can be a test-mass limit for a two-body source.}
\begin{equation}
        \boxed{
        x(t)=\eps\frac{c^2}{\kappa}\sqrt{\W\!\left(\ee^{\kappa t/c}\right)}
        }.
        \label{eq:trajectory}
\end{equation} 
The constant $\kappa$ serves as an acceleration scale, and $\eps\ll1$ is a dimensionless parameter that controls the nonrelativistic approximation. 

The source ingredient $q(t)$ is given by 
\begin{equation}
        \q(t)=x(t)^2=\eps^2\frac{c^4}{\kappa^2}\,y(t).
        \label{eq:qdef}
\end{equation}
Eq.~\eqref{eq:inverse1} then identifies a kinematic invariant, 
\begin{equation}
        \boxed{
        c\frac{\dd}{\dd t}
        \left[
        y(t)
        +
        \ln y(t)
        \right]
        =\kappa
        } .
        \label{eq:peel}
\end{equation}
This is reminiscent of the kinematic invariant associated with the acceleration history of a moving mirror called the peel acceleration, see e.g.\  \cite{Barcelo:2010pj, Barcelo:2010xk, Bianchi:2014qua, Bianchi:2014vea, carlitz1987reflections, CW2lifetime,good2013time,Ievlev:2023xzv}.  In the black hole case, the square of the peeling function is the power emitted, which resolves the problem of negative energy flux; see \cite{Bianchi:2026xoi}. In the thermal Schwarzschild case \cite{Good:2016oey}, $\kappa$ plays the role of the surface gravity, $\kappa = c^4/4GM$.  In our case, the invariant \(\kappa\) sets the scale of the system, but is also the constant logarithmic-linear growth rate of the dimensionless quadrupole coordinate, Eq.~(\ref{eq:peel}). 

The source is obtained from
\begin{equation}
        \frac{\dd y}{\dd u}=\frac{y}{1+y},\qquad
        \frac{\dd^3 y}{\dd u^3}=\frac{y(1-2y)}{(1+y)^5}.
        \label{eq:y_derivatives}
\end{equation}
Using Eq.~\eqref{eq:qdef},
\begin{equation}
        \boxed{
        \St(t)=\frac{\dd^3\q}{\dd t^3}
        =
        \eps^2 c\kappa\,
        \frac{y(1-2y)}{(1+y)^5}
        } .
        \label{eq:source}
\end{equation}
The velocity is
\begin{equation}
        v(t)=\dot x(t)
        =
        \frac{\eps c}{2}
        \frac{\sqrt y}{1+y},
        \label{eq:velocity}
\end{equation}
with a maximum
\begin{equation}
        \frac{v_{\rm max}}{c}=\frac{\eps}{4},
        \label{eq:vmax}
\end{equation}
which occurs at $y=1$.  Hence $\eps\ll1$ is the nonrelativistic control parameter.  The timelike motion starts at rest (at $y=0$, $t=-\infty$) and returns asymptotically to rest 
(at $y=+\infty$, $t=+\infty$),
\begin{equation}
        v(-\infty)=0,\qquad v(+\infty)=0,
        \label{eq:rest_to_rest}
\end{equation}
although its displacement is unbounded: $x(+\infty)=+\infty$.  
Figure~\ref{fig:penroseGrav} illustrates the trajectory; notice 
the absence of a horizon.

\section{Spectrum, energy, and graviton count}
\label{sec:spectrum}

For the trajectory in Eq.~\eqref{eq:trajectory} 
the source Fourier transform to compute the radiation spectrum is
\begin{equation}
        \widehat{\St}(\omega)
        =
        \frac{\eps^2 c^2}{\sqrt{2\pi}}
        \int_{-\infty}^{\infty}
        e^{-i\Omega u}
        \frac{\dd^3}{\dd u^3}\W(e^u)\,\dd u .
        \label{eq:source_transform}
\end{equation}
The relevant Fourier integral entering Eq.~\eqref{eq:spectrumgeneral}, in the same symmetric Fourier normalization convention, with $\Omega > 0$, is
\begin{equation}
        \left|
        \frac{1}{\sqrt{2\pi}}
        \int_{-\infty}^{\infty}
        e^{-i\Omega u}
        \frac{\dd^3}{\dd u^3}\W(e^u)\,\dd u
        \right|^2
        =
        \frac{\Omega^3}{e^{2\pi\Omega}-1}.
        \label{eq:dimensionless_identity}
\end{equation} 
Therefore the energy spectrum is
\begin{equation}
        \boxed{
        \frac{\dd E}{\dd\omega}
        =
        \frac{4Gm^2 c^2\eps^4}{15\kappa^3}
        \frac{\omega^3}{e^{2\pi c\omega/\kappa}-1}
        } .
        \label{eq:planckspectrum}
\end{equation}
This is the Planck energy spectrum.  
Equivalently, writing $2\pi c\omega/\kappa =\hbar\omega/(k_B T)$, the temperature in Kelvin is
\begin{equation}
                T=\frac{\hbar\kappa}{2\pi c k_B}.
        \label{eq:temperature}
\end{equation}

The total emitted energy is finite:
\begin{align}
        E
        &=
        \int_{0}^{\infty}
        \frac{\dd E}{\dd\omega}\,\dd\omega                 \nonumber\\
        &=
        \frac{4Gm^2 c^2\eps^4}{15\kappa^3}
        \int_{0}^{\infty}
        \frac{\omega^3}{e^{2\pi c\omega/\kappa}-1}\,\dd\omega .
        \label{eq:energy_integral}
\end{align}
Using
\begin{equation}
        \int_{0}^{\infty}
        \frac{\omega^3}{e^{2\pi c\omega/\kappa}-1}\,\dd\omega
        =
        \frac{\kappa^4}{240c^4},
        \label{eq:omega_integral}
\end{equation}
we obtain
\begin{equation}
        \boxed{
        E=\frac{Gm^2\eps^4\kappa}{900c^2}
        } .
        \label{eq:totalenergy}
\end{equation} 
Interestingly, the linear scaling \(E\propto\kappa\), or equivalently \(E\propto T\), is not in
conflict with the Stefan--Boltzmann law.  Eq.~\eqref{eq:planckspectrum}
is the total emitted-energy spectrum of a prescribed quadrupole history, not an
equilibrium radiance per unit area.  The explicit \(\kappa^{-3}\) normalization in the spectrum cancels three of
the four powers generated by the Planck integral.  The Planck factor, therefore,
fixes the frequency dependence, while the overall scaling is set by the
kinematic quadrupole source rather than by an emitting surface or an
equilibrium density of states. 

The same value follows directly from the time-domain power.  From Eq.~\eqref{eq:source},
\begin{align}
        \int_{-\infty}^{\infty}\St(t)^2\,\dd t
        &=
        \eps^4 c^3\kappa
        \int_{0}^{\infty}
        \frac{y(1-2y)^2}{(1+y)^9}\,\dd y                 \nonumber\\
        &=
        \frac{\eps^4 c^3\kappa}{120}.
        \label{eq:stime}
\end{align}
Therefore
\begin{equation}
        \int_{-\infty}^{\infty}P(t)\,\dd t
        =
        \frac{2Gm^2}{15c^5}
        \frac{\eps^4 c^3\kappa}{120}
        =
        \frac{Gm^2\eps^4\kappa}{900c^2},
        \label{eq:etime}
\end{equation}
in agreement with Eq.~\eqref{eq:totalenergy}.

We define the particle number\footnote{In perturbative quantum gravity, gravitational radiation is described in terms
of massless spin--2 graviton modes~\cite{Weinberg:1965rz,Weinberg:1965nx}.
For a classical external stress--energy source, Laga and Suyama~\cite{Laga:2026vwm} showed that the
expectation value of the quantum gravitational-wave operator reproduces the
retarded classical waveform, and that the corresponding emitted-graviton
statistics are Poissonian, as expected for a coherent state.
Following this quantum-field interpretation of classical-source radiation (see, e.g., the semiclassical photon analog in Jackson \cite{Jackson:490457}), we
assign the energy \(\hbar\omega\) to each emitted graviton.} 
spectrum by dividing the classical energy spectrum Eq.~(\ref{eq:planckspectrum}) by \(\hbar\omega\):

\begin{equation}
        \boxed{
        \frac{\dd N}{\dd\omega}
        =
        \frac{4Gm^2 c^2\eps^4}{15\hbar\kappa^3}
        \frac{\omega^2}{e^{2\pi c\omega/\kappa}-1}
        } .
        \label{eq:numberspectrum}
\end{equation}
The total graviton count is finite,
\begin{align}
        N
        &=
        \frac{4Gm^2 c^2\eps^4}{15\hbar\kappa^3}
        \int_0^\infty
        \frac{\omega^2}{e^{2\pi c\omega/\kappa}-1}\,\dd\omega       \nonumber\\
        &=
        \boxed{
        \frac{Gm^2\eps^4\zeta(3)}{15\pi^3\hbar c}
        } .
        \label{eq:totalnumber}
\end{align}
Thus, the total graviton number is independent of the temperature scale $\kappa = 2\pi c k_B T/\hbar$. Both the total energy and the characteristic graviton energy are proportional to the kinematic invariant \(\kappa\). 

The finite value of \(N\), Eq.~(\ref{eq:totalnumber}), is the average graviton occupation associated with the emitted spectrum, Eq.~(\ref{eq:planckspectrum}). It should not be confused with a detector-count proposal.\footnote{Following Dyson's classic discussion of graviton detectability~\cite{Dyson:2013hbl}, recent work has proposed quantum-sensing approaches to single-graviton detection~\cite{Tobar:2023single} although emphasizing that single-graviton detector clicks alone need not rule out classical gravitational-wave description~\cite{Carney:2023graviton}.}  

\section{Deformations of the Log Clock}
\label{sec:deformations}

The product-log trajectory in Eq.~\eqref{eq:trajectory}
is the minimal way to combine two requirements.  The
logarithm supplies the imaginary-time branch structure,
while the added linear term supplies the late-time decay
needed for an ordinary Fourier spectrum.  It is useful to
separate these two roles by replacing the linear repair
with a positive-power repair,
\begin{equation}
        u=\ln y+y^p,
        \qquad p>0 .
        \label{eq:deformed_clock}
\end{equation}
The case \(p=1\) is the product-log clock used above.
For integer \(p\), the repair term \(y^p\) is single-valued
around \(y=0\), so the only monodromy is still
\(\ln y\to \ln y+2\pi i\).  Noninteger values of \(p\) are also useful as continuous spectral deformations, although in that case, the repair term itself introduces additional
branch structure.

\begin{figure}[t]
\centering
\includegraphics[width=0.48\textwidth,height=0.40\textheight,keepaspectratio]{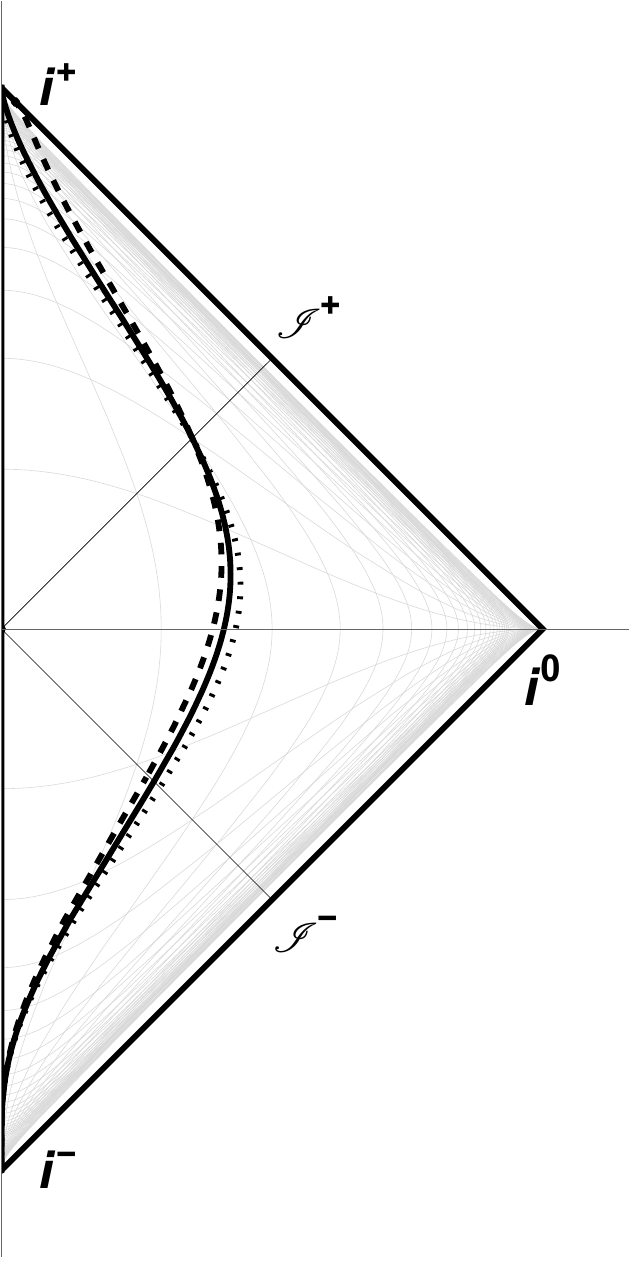}
\caption{
Penrose conformal compactification of the rectilinear source trajectories, Eq.~(\ref{eq:xp_deformed}), associated with the Lambert-$W$ family of Planckian graviton emitters. The dashed, solid, and dotted curves correspond to $p=\tfrac12$, $p=1$, and $p=2$, respectively. The central trajectory ($p=1$) yields the exact Planckian graviton spectrum, while the neighboring cases introduce deviations from exact thermality. Note how the $p=1/2$ case has asymptotic constant future velocity rather than coming to rest. 
For illustration here $\epsilon=\kappa = c = 1$, giving $v_\textrm{max} = 1/4$. 
}
\label{fig:penroseGrav}
\end{figure}

Eq.~\eqref{eq:deformed_clock} can be solved in
terms of the Lambert function.  Writing \(z=y^p\), one
has
\begin{equation}
        p u=\ln z+pz ,
\end{equation}
and hence
\begin{equation}
        y_p(u)
        =
        \left[
        \frac{1}{p}\,
        \W\!\left(p\,\ee^{p u}\right)
        \right]^{1/p}.
        \label{eq:yp_solution}
\end{equation}
Thus, the corresponding rectilinear trajectory is
\begin{equation}
        x_p(t)
        =
        \eps\frac{c^2}{\kappa}\sqrt{y_p(t)},
        \qquad
        \q_p(t)=x_p(t)^2
        =
        \eps^2\frac{c^4}{\kappa^2}\,y_p(t).
        \label{eq:xp_deformed}
\end{equation} 
The original trajectory is recovered for 
\(p=1\), since \(y_1(u)=\W(\ee^u)\). See Figure \ref{fig:penroseGrav} for illustration.

Differentiating Eq.~\eqref{eq:deformed_clock} gives
\begin{equation}
        \frac{\dd y}{\dd u}
        =
        \frac{y}{1+p y^p}.
        \label{eq:deformed_first_derivative}
\end{equation}
Therefore, the velocity is
\begin{equation}
        v_p(t)=\dot x_p(t)
        =
        \frac{\eps c}{2}\,
        \frac{\sqrt{y}}{1+p y^p}.
        \label{eq:deformed_velocity}
\end{equation}
This makes the role of the repair term transparent.  The
factor \(\sqrt y\) is inherited from the logarithmic clock,
while the denominator regulates the late-time motion.  At
large \(y\),
\begin{equation}
        \frac{v_p}{c}
        \sim
        \frac{\eps}{2p}\,y^{1/2-p}.
        \label{eq:deformed_velocity_asymptotic}
\end{equation}
Thus, the source returns asymptotically to rest for
\(p>1/2\), approaches a constant velocity for \(p=1/2\),
and eventually leaves the rest-to-rest nonrelativistic
class for \(p<1/2\). 
(One could also turn this derivation around by starting with a velocity regularization of the form Eq.~(\ref{eq:deformed_velocity}) and deriving Eq.~(\ref{eq:deformed_clock}) from it.) 

The dimensionless quadrupole source is obtained by three
derivatives with respect to \(u\):
\begin{equation}
        \frac{\dd^3 y}{\dd u^3}
        =
        \frac{
        y\left[
        1
        -
        p(p^2+3p-2)y^p
        +
        p^2(p-1)(2p-1)y^{2p}
        \right]
        }
        {(1+p y^p)^5}.
        \label{eq:deformed_source_dimensionless}
\end{equation}
The physical source is therefore
\begin{equation}
        \St_p(t)
        \equiv
        \frac{\dd^3\q_p}{\dd t^3}
        =
        \eps^2 c\kappa\,
        \frac{\dd^3 y}{\dd u^3}.
        \label{eq:deformed_source_physical}
\end{equation}
As \(u\to-\infty\), one has \(y\sim \ee^u\), and hence
\begin{equation}
        \frac{\dd^3 y}{\dd u^3}\sim \ee^u .
        \label{eq:deformed_past}
\end{equation}
The early-time behavior is therefore unchanged: all
members of the family retain the logarithmic clock near
the branch point.  At late times, \(y^p\sim u\), so
\(y\sim u^{1/p}\).  For generic \(p\neq1,1/2\),
\begin{equation}
        \frac{\dd^3 y}{\dd u^3}
        \sim
        \frac{(p-1)(2p-1)}{p^3}\,
        y^{1-3p}
        \sim
        \frac{(p-1)(2p-1)}{p^3}\,
        u^{1/p-3} \ ,
        \label{eq:deformed_future_generic}
\end{equation} 
and thus $y'''$ vanishes at late times as $u\to +\infty$ for $p>1/3$. 
The coefficient of this leading term vanishes at
\(p=1\) and \(p=1/2\).  In particular, for the product-log
case \(p=1\),
\begin{equation}
        \frac{\dd^3 y}{\dd u^3}
        =
        \frac{y(1-2y)}{(1+y)^5}
        \sim
        -\frac{2}{u^3},
        \qquad u\to+\infty,
        \label{eq:productlog_late_tail}
\end{equation}
which is the late-time falloff used in the exact 3D
Planckian construction.

The corresponding Fourier transform can also be written
in closed form.  Define
\begin{equation}
        {\cal I}_p(\Omega)
        =
        \frac{1}{\sqrt{2\pi}}
        \int_{-\infty}^{\infty}
        \ee^{-\ii\Omega u}
        \frac{\dd^3 y}{\dd u^3}\,\dd u,
        \qquad
        \Omega=\frac{c\omega}{\kappa}.
        \label{eq:Ip_definition}
\end{equation}
For \(p>1/2\), the source has sufficient late-time decay
for the transform to be understood as an ordinary
oscillatory integral, with the usual small damping
prescription.  Integrating twice by parts gives
\begin{align}
        {\cal I}_p(\Omega)
        &=
        \frac{(\ii\Omega)^2}{\sqrt{2\pi}}
        \int_{-\infty}^{\infty}
        \ee^{-\ii\Omega u}
        \frac{\dd y}{\dd u}\,\dd u                         \nonumber\\
        &=
        \frac{(\ii\Omega)^2}{\sqrt{2\pi}}
        \int_{0}^{\infty}
        y^{-\ii\Omega}
        \ee^{-\ii\Omega y^p}\,\dd y                         \nonumber\\
        &=
        \frac{(\ii\Omega)^2}{p\sqrt{2\pi}}\,
        (\ii\Omega)^{-(1-\ii\Omega)/p}
        \Gamma\!\left(\frac{1-\ii\Omega}{p}\right).
        \label{eq:Ip_closed}
\end{align}
Therefore
\begin{equation}
        |{\cal I}_p(\Omega)|^2
        =
        \frac{\Omega^{4-2/p}}{2\pi p^2}\,
        \ee^{-\pi\Omega/p}
        \left|
        \Gamma\!\left(\frac{1-\ii\Omega}{p}\right)
        \right|^2 .
        \label{eq:Ip_modulus}
\end{equation}
The energy spectrum of the deformed clock is then
\begin{equation}
        \frac{\dd E_p}{\dd\omega}
        =
        \frac{4Gm^2\eps^4}{15c}\,
        |{\cal I}_p(c\omega/\kappa)|^2 .
        \label{eq:deformed_energy_spectrum}
\end{equation}

Eq.~\eqref{eq:Ip_modulus} makes clear why the
product-log clock \(p=1\) is distinguished within the
deformed family.  For \(p=1\),
\begin{equation}
        |{\cal I}_1(\Omega)|^2
        =
        \frac{\Omega^2}{2\pi}\,
        \ee^{-\pi\Omega}
        |\Gamma(1-\ii\Omega)|^2 .
        \label{eq:I1_before_planck}
\end{equation}
Using
\begin{equation}
        |\Gamma(1-\ii\Omega)|^2
        =
        \frac{\pi\Omega}{\sinh\pi\Omega},
        \label{eq:gamma_identity_productlog}
\end{equation}
one obtains
\begin{equation}
        |{\cal I}_1(\Omega)|^2
        =
        \frac{\Omega^3}{\ee^{2\pi\Omega}-1}.
        \label{eq:I1_planck}
\end{equation}
Substitution into Eq.~\eqref{eq:deformed_energy_spectrum}
reproduces Eq.~\eqref{eq:planckspectrum}.  Thus the
linear repair \(y\) is a convenient
regularization and the member of this family for which the logarithmic clock and the Fourier transform combine
to give the exact 3D Planck energy spectrum.

For \(p\neq1\), the logarithmic branch still organizes the
analytic structure, but the spectral weight is deformed by
the gamma function in Eq.~\eqref{eq:Ip_modulus}.  In the
infrared,
\begin{equation}
        |{\cal I}_p(\Omega)|^2
        \sim
        \frac{\Gamma(1/p)^2}{2\pi p^2}\,
        \Omega^{4-2/p},
        \qquad
        \Omega\to0 .
        \label{eq:Ip_infrared}
\end{equation}
Hence
\begin{equation}
        \frac{\dd E_p}{\dd\omega}
        \propto
        \omega^{4-2/p},
        \qquad
        \frac{\dd N_p}{\dd\omega}
        \propto
        \omega^{3-2/p},
        \qquad
        \omega\to0 .
        \label{eq:deformed_ir_scaling}
\end{equation}
The total radiated energy is infrared finite for
\(p>2/5\), while the total graviton number is infrared
finite for \(p>1/2\).  The latter condition is the same
condition that makes the velocity in
Eq.~\eqref{eq:deformed_velocity} return to zero at late
times.  Thus the physically clean rest-to-rest,
finite-graviton deformations satisfy \(p>1/2\), with the
product-log clock \(p=1\) sitting safely inside this class.

This also clarifies why the late-time tail alone does not
determine the spectrum in the low-frequency limit.  In the product-log
case the source behaves as \(\dd^3y/\dd u^3\sim -2/u^3\)
at late times, but the exact transform gives
\begin{equation}
        |{\cal I}_1(\Omega)|^2
        =
        \frac{\Omega^3}{\ee^{2\pi\Omega}-1}
        \sim
        \frac{\Omega^2}{2\pi},
        \qquad
        \Omega\to0 .
        \label{eq:productlog_ir}
\end{equation}
The infrared behavior is controlled by the full source
history and its global moment structure.  The deformed family, therefore, 
teaches a useful lesson: 
the logarithm supplies the imaginary-time clock, the
positive-power term supplies convergence, and the
linear repair \(p=1\) is the special choice that yields the
exact Planckian graviton spectrum.

\section{Discussion}
\label{sec:discussion}

The preceding section shows that the exact Planck form is
not a generic consequence of adding any late-time repair to
the logarithmic clock; it is selected by the product-log
member \(p=1\).  The result is best interpreted
as a kinematic out-of-equilibrium construction.  No horizon is present.  The Planck factor arises because the dimensionless quadrupole coordinate satisfies the inverse relation $y+\ln y=u$, whose logarithmic branch structure for imaginary-time periodicity produces the Bose-Einstein factor in frequency space (as for the black hole case).  

We emphasize the contrast with electromagnetic radiation induced by the acceleration of a charge.  
For rectilinear motion, this distinction is illustrated:
\begin{equation}
        \textrm{photons:}\quad \ddot x(t),
        \qquad
        \textrm{gravitons:}\quad \frac{\dd^3}{\dd t^3}x(t)^2 .
        \label{eq:photons_gravitons}
\end{equation} 
Nonrelativistic gravitational radiation is governed by the third derivative of the mass quadrupole.  
Thus, Eq.~\eqref{eq:trajectory} should not be read as making the acceleration
\(\ddot x(t)\) Planckian.  A (timelike) worldline \(x(t)\) of course enters both cases, but the radiating source functional is different: electromagnetic
dipole radiation involves \(\ddot x(t)\), whereas the present gravitational
calculation involves \(\dd^3 x(t)^2/\dd t^3\).  It is this quadrupole source,
not the bare acceleration, whose Fourier transform gives the Planck factor.

\subsection{Normalized quadrupole source primitive}
\label{subsec:quadrupole-source-primitive}

The same comparison suggests a useful dimensionless primitive for the
radiating source, analogous to the von Neumann entanglement entropy for particle production in contexts like mirrors, black holes, and beta decay \cite{Good:2025qta}. 

Since the gravitational power is controlled by the
third derivative of the quadrupole coordinate \(q=x^2\), rather than by the acceleration \(\ddot x\), define
\begin{equation}
        {\cal S}_g(t)
        \equiv
        \sqrt{\frac{\pi}{45}}\,
        \frac{1}{c^2}\frac{\dd^2}{\dd t^2}x(t)^2 .
        \label{eq:Sg_def}
\end{equation}
This quantity is dimensionless. 
Recall that when the power is sourced by the acceleration, the entanglement entropy ${\cal S}_{\rm acc}\sim \dot x$, i.e.,\ one less time derivative than the source, so our definition ${\cal S}_g$ is directly analogous.

For the product-log clock
\(y+\ln y=u\), with \(q=\eps^2 c^4 y/\kappa^2\), it becomes
\begin{equation}
        {\cal S}_g(u)
        =
        \eps^2\sqrt{\frac{\pi}{45}}\,
        \frac{y}{(1+y)^3}.
        \label{eq:Sg_y}
\end{equation}
Thus \({\cal S}_g\) vanishes as \(u\to\pm\infty\) and has the finite
maximum
\begin{equation}
        {\cal S}_{g,\max}
        =
        \frac{4}{27}\,
        \eps^2\sqrt{\frac{\pi}{45}},
        \qquad y=\frac{1}{2}.
        \label{eq:Sg_max}
\end{equation}
It is therefore a localized quadrupole-source amplitude associated with
the prescribed history.

The normalization in Eq.~\eqref{eq:Sg_def} is useful because its time derivative is exactly proportional to the radiating source:
\begin{equation}
        \dot{\cal S}_g
        =
        \sqrt{\frac{\pi}{45}}\,
        \frac{1}{c^2}
        \frac{\dd^3}{\dd t^3}x(t)^2 .
        \label{eq:Sg_dot}
\end{equation}
Consequently, the instantaneous power may be written as
\begin{equation}
        P(t)
        =
        \frac{6Gm^2}{\pi c}\,
        \dot{\cal S}_g^{\,2},
        \label{eq:power_Sg}
\end{equation} 
in close analogy to the acceleration case, i.e.,\ the Larmor formula $P(t)\sim \dot{\cal S}_{\rm acc}^{\,2}$. 
Equivalently,
\begin{equation}
        E
        =
        \frac{6Gm^2}{\pi c}
        \int_{-\infty}^{\infty}
        \dot{\cal S}_g^{\,2}\,\dd t
        =
        \frac{Gm^2\eps^4\kappa}{900c^2},
        \label{eq:energy_Sg}
\end{equation}
which is the same total energy obtained from the Planckian frequency
spectrum.  This gives a compact consistency check: the entropic variable whose derivative gives the quadrupole radiation source also reproduces the time-domain norm of the emitted radiation.

While the entropy object Eq.~\eqref{eq:Sg_def} is not the equilibrium thermodynamic entropy, it plays an operational role as the normalized primitive of the quadrupole source.  In the full tensor description, the
corresponding object is therefore additive in the mass density because \(I_{ij}[\rho_1+\rho_2]=I_{ij}[\rho_1]+I_{ij}[\rho_2]\); thus \(\mathcal S_{ij}\propto \ddot I_{ij}\). In this sense \({\cal S}_g\) packages
the relevant source information: its derivative carries the radiated power, and the logarithmic clock \(y+\ln y=u\) supplies the analytic structure responsible for the spectral Bose-Einstein factor.

\subsection{Frequency spectrum versus angular pattern}
The Planckian character of Eq.~(\ref{eq:planckspectrum}) is a statement about the
frequency spectrum, without an angular distribution. The radiation field is not isotropic.
Rectilinearity selects a preferred spatial axis, and the transverse-traceless projection
of the corresponding trace-free quadrupole gives the usual spin-two angular pattern.
If $\theta$ denotes the angle between the observation direction and the rectilinear axis, then
\begin{equation}
\frac{dE}{d\omega\,d\Omega_{\hat{\mathbf n}}}
=
\frac{15}{32\pi}\sin^4\theta\,\frac{dE}{d\omega}.
\label{eq:angular-spectrum}
\end{equation}
Here $d\Omega_{\hat{\mathbf n}}=\sin\theta\,d\theta\,d\phi$ is the solid-angle element on the observation sphere, with
\(0\leq\theta\leq\pi\) and \(0\leq\phi<2\pi\). The factor $15/(32\pi)$
normalizes the angular distribution, since
$\int d\Omega_{\hat{\mathbf n}}\,\sin^4\theta=32\pi/15$.

A strictly isotropic breathing mode does not produce the present radiation because a
spherically symmetric expansion or contraction changes only the monopole part of the
source; its trace-free quadrupole moment vanishes, and hence there is no gravitational
radiation at this order. Thus, while the logarithmic imaginary-time structure provides the directionally independent Bose-Einstein factor, the spin-two quadrupole projection provides the angular dependence. The radiation distribution is therefore Planckian in frequency but anisotropic in angle,
and the thermal signature is out of equilibrium in the absence of a blackbody field.

\section{Conclusions}
\label{sec:concl}
We have shown that the standard quadrupole formula reduces, for rectilinear nonrelativistic motion, to a 
source term 
proportional to $[\dddot{x^2}]^2$ where the product-log trajectory 
\begin{equation}
        x(t)=\eps\frac{c^2}{\kappa}\sqrt{W\!\left(e^{\kappa t/c}\right)},
        \label{eq:conclusion_trajectory}
\end{equation}
gives an exact Planck spectrum for the emitted gravitational radiation,
\begin{equation}
        \frac{\dd E}{\dd\omega}
        =
        \frac{4Gm^2 c^2\eps^4}{15\kappa^3}
        \frac{\omega^3}{e^{2\pi c\omega/\kappa}-1}.
        \label{eq:conclusion_spectrum}
\end{equation}
The total energy and total graviton number are
\begin{equation}
        E=\frac{Gm^2\eps^4\kappa}{900c^2},
        \qquad
        N=\frac{Gm^2\eps^4\zeta(3)}{15\pi^3\hbar c}.
        \label{eq:conclusion_totals}
\end{equation}
The construction demonstrates exact Planckian graviton radiation from nonrelativistic quadrupole kinematics.  Its physical significance demonstrates both a thermal clock in imaginary time and a spectral origin rather than a thermodynamic one: the radiation is Planck-distributed, without an equilibrium bath. 

There is no spacetime horizon in this construction.  The variables
\(u=\kappa t/c\) and \(y\propto x^2\) range over
\(u\in(-\infty,\infty)\) and \(y\in(0,\infty)\), with the displacement
unbounded only asymptotically at late times.  Thus, the Planck factor is not attributed to a causal boundary, a null-shell geometry, or a statistical KMS periodicity, but to the logarithmic monodromy of the prescribed quadrupole source history.

\section*{Acknowledgements}
Funding comes partly from the FY2024-SGP-1-STMM Faculty Development Competitive Research Grant (FDCRGP) no.201223FD8824 and SSH20224004 at Nazarbayev University in Qazaqstan. Appreciation is given to the ROC (Taiwan) Ministry of Science and Technology (MOST), Grant no.112-2112-M-002-013, National Center for Theoretical Sciences (NCTS), and Leung Center for Cosmology and Particle Astrophysics (LeCosPA) of National Taiwan University.


\bibliography{main} 

@article{Carney:2023graviton,
    author = "Carney, Daniel and Domcke, Valerie and Rodd, Nicholas L.",
    title = "{Graviton detection and the quantization of gravity}",
    eprint = "2308.12988",
    archivePrefix = "arXiv",
    primaryClass = "hep-th",
    reportNumber = "CERN-TH-2023-155",
    doi = "10.1103/PhysRevD.109.044009",
    journal = "Phys. Rev. D",
    volume = "109",
    number = "4",
    pages = "044009",
    year = "2024"
}

@article{Tobar:2023single,
    author = "Tobar, Germain and Manikandan, Sreenath K. and Beitel, Thomas and Pikovski, Igor",
    title = "{Detecting single gravitons with quantum sensing}",
    eprint = "2308.15440",
    archivePrefix = "arXiv",
    primaryClass = "quant-ph",
    reportNumber = "NORDITA 2023-040",
    doi = "10.1038/s41467-024-51420-8",
    journal = "Nature Commun.",
    volume = "15",
    number = "1",
    pages = "7229",
    year = "2024"
}

@article{Dyson:2013hbl,
    author = "Dyson, Freeman",
    title = "{Is a graviton detectable?}",
    doi = "10.1142/S0217751X1330041X",
    journal = "Int. J. Mod. Phys. A",
    volume = "28",
    pages = "1330041",
    year = "2013"
}

@article{Laga:2026vwm,
    author = "Laga, Felix and Suyama, Teruaki",
    title = "{Quantum description of gravitational waves generated by a classical source}",
    eprint = "2604.20228",
    archivePrefix = "arXiv",
    primaryClass = "gr-qc",
    month = "4",
    journal = {},
    year = "2026"
}

@article{Weinberg:1965rz,
    author = "Weinberg, Steven",
    title = "{Photons and gravitons in perturbation theory: Derivation of Maxwell's and Einstein's equations}",
    journal = "Phys. Rev.",
    volume = "138",
    pages = "B988--B1002",
    year = "1965",
    doi = "10.1103/PhysRev.138.B988",
    url = "https://doi.org/10.1103/PhysRev.138.B988"
}

@article{Weinberg:1965nx,
    author = "Weinberg, Steven",
    title = "{Infrared photons and gravitons}",
    journal = "Phys. Rev.",
    volume = "140",
    pages = "B516--B524",
    year = "1965",
    doi = "10.1103/PhysRev.140.B516",
    url = "https://doi.org/10.1103/PhysRev.140.B516"
}

@article{Bianchi:2026xoi,
    author = "Bianchi, Eugenio and Paraizo, Daniel E.",
    title = "{On Radiative Fluxes and Coulombic Charges in the Balance Law for Black Hole Evaporation}",
    eprint = "2603.13120",
    archivePrefix = "arXiv",
    primaryClass = "gr-qc",
    month = "3",
    journal = {},
    year = "2026"
}

@article{Good:2025qta,
    author = "Good, Michael R. R. and Ievlev, Evgenii and Linder, Eric V.",
    title = "{Particle creation from entanglement entropy}",
    eprint = "2508.17067",
    archivePrefix = "arXiv",
    primaryClass = "quant-ph",
    reportNumber = "FTPI-MINN-25-10, UMN-TH-4504/25",
    doi = "10.1093/ptep/ptaf183",
    journal = "PTEP",
    volume = "2026",
    pages = "1",
    year = "2026"
}

@article{Gorban:2024vss,
    author = "Gorban, Matthew J. and Julius, William D. and Brown, Patrick M. and Matulevich, Jacob A. and Radhakrishnan, Ramesh and Cleaver, Gerald B.",
    title = "{First- and Second-Order Forces in the Asymmetric Dynamical Casimir Effect for a Single \ensuremath{\delta}-\ensuremath{\delta}' Mirror}",
    doi = "10.3390/physics6020047",
    journal = "MDPI Physics",
    volume = "6",
    number = "2",
    pages = "760--779",
    year = "2024"
}

@article{Good:2022eub,
    author = "Good, Michael R. R. and Davies, Paul C. W.",
    title = "{Infrared acceleration radiation}",
    eprint = "2206.07291",
    archivePrefix = "arXiv",
    primaryClass = "gr-qc",
    doi = "10.1007/s10701-023-00694-x",
    journal = "Found. Phys.",
    volume = "53",
    number = "3",
    pages = "53",
    year = "2023"
}

@article{Agullo:2025opy,
    author = "Agullo, Ivan and Calizaya Cabrera, Paula and Navascu{\'e}s, Beatriz Elizaga",
    title = "{Vacuum-purified Hawking radiation from evaporating black holes: Lessons from moving mirrors}",
    eprint = "2512.18354",
    archivePrefix = "arXiv",
    primaryClass = "gr-qc",
     journal = "",
    month = "12",
    year = "2025"
}

@article{Kumar:2023kse,
    author = "Kumar, Piyush and Reyes, Ignacio A. and Wintergerst, Jakob",
    title = "{Relativistic dynamics of moving mirrors in CFT2: Quantum backreaction and black holes}",
    eprint = "2310.03483",
    archivePrefix = "arXiv",
    primaryClass = "hep-th",
    doi = "10.1103/PhysRevD.109.065010",
    journal = "Phys. Rev. D",
    volume = "109",
    number = "6",
    pages = "065010",
    year = "2024"
}

@article{Reyes:2021npy,
    author = "Reyes, Ignacio A.",
    title = "{Moving mirrors, page curves, and bulk entropies in AdS2}",
    eprint = "2103.01230",
    archivePrefix = "arXiv",
    primaryClass = "hep-th",
    doi = "10.1103/PhysRevLett.127.051602",
    journal = "Phys. Rev. Lett.",
    volume = "127",
    number = "5",
    pages = "051602",
    year = "2021"
}

@article{Zhakenuly:2021pfm,
    author = "Zhakenuly, Abay and Temirkhan, Maksat and Good, Michael R. R. and Chen, Pisin",
    title = "{Quantum power distribution of relativistic acceleration radiation: classical electrodynamic analogies with perfectly reflecting moving mirrors}",
    eprint = "2101.02511",
    archivePrefix = "arXiv",
    primaryClass = "gr-qc",
    doi = "10.3390/sym13040653",
    journal = "Symmetry",
    volume = "13",
    number = "4",
    pages = "653",
    year = "2021"
}

@article{Hawking:1974rvNATURE,
    author = "Hawking, S. W.",
    title = "{Black hole explosions}",
    doi = "10.1038/248030a0",
    journal = "Nature",
    volume = "248",
    pages = "30--31",
    year = "1974"
}

@article{Ievlev:2023inj,
    author = "Ievlev, Evgenii and Good, Michael R. R.",
    title = "{Thermal Larmor Radiation}",
    eprint = "2303.03676",
    archivePrefix = "arXiv",
    primaryClass = "gr-qc",
    doi = "10.1093/ptep/ptae042",
    journal = "PTEP",
    volume = "2024",
    number = "4",
    pages = "043A01",
    year = "2024"
}

@article{Ievlev:2023xzv,
    author = "Ievlev, Evgenii and Good, Michael R. R. and Linder, Eric V.",
    title = "{IR-finite thermal acceleration radiation}",
    eprint = "2304.04412",
    archivePrefix = "arXiv",
    primaryClass = "gr-qc",
    doi = "10.1016/j.aop.2024.169593",
    journal = "Annals Phys.",
    volume = "461",
    pages = "169593",
    year = "2024"
}

@article{Lynch:2022rqx,
    author = "Lynch, Morgan H. and Ievlev, Evgenii and Good, Michael R. R.",
    title = "{Accelerated electron thermometer: observation of 1D Planck radiation}",
    eprint = "2211.14774",
    archivePrefix = "arXiv",
    primaryClass = "nucl-ex",
    doi = "10.1093/ptep/ptad157",
    journal = "PTEP",
    volume = "2024",
    number = "2",
    pages = "023D01",
    year = "2024"
}

@article{CW2lifetime,
  title = {Lifetime of a black hole},
  author = {Carlitz, Robert D. and Willey, Raymond S.},
  journal = {Phys. Rev. D},
  volume = {36},
  issue = {8},
  pages = {2336--2341},
  numpages = {0},
  year = {1987},
  month = {Oct},
  publisher = {American Physical Society},
  doi = {10.1103/PhysRevD.36.2336},
  url = {https://link.aps.org/doi/10.1103/PhysRevD.36.2336}
}

@article{DeWitt:1975ys,
    author = "DeWitt, Bryce S.",
    title = "{Quantum field theory in curved space-time}",
    doi = "10.1016/0370-1573(75)90051-4",
    journal = "Phys. Rept.",
    volume = "19",
    pages = "295--357",
    year = "1975"
}

@article{Davies:1977yv,
    author = "Davies, P.C.W. and Fulling, S.A.",
    doi = "10.1098/rspa.1977.0130",
    journal = "Proc. R. Soc. Lond. A",
    pages = "237--257",
    title = "{Radiation from moving mirrors and from black holes}",
    volume = "356",
    year = "1977"
}

@article{Davies:1976hi,
author = {Fulling, S. A.  and Davies, P. C. W.},
title = "Radiation from a Moving Mirror in Two Dimensional Space-Time: Conformal Anomaly",
journal = {Proc. R. Soc. Lond. A},
volume = {348},
number = {1654},
pages = {393-414},
year = {1976},
URL = {https://royalsocietypublishing.org/doi/abs/10.1098/rspa.1976.0045}
}

@article{Chen:2015bcg,
    author = "Chen, Pisin and Mourou, Gerard",
    archivePrefix = "arXiv",
    doi = "10.1103/PhysRevLett.118.045001",
    eprint = "1512.04064",
    journal = "Phys. Rev. Lett.",
    number = "4",
    pages = "045001",
    primaryClass = "gr-qc",
    title = "{Accelerating Plasma Mirrors to Investigate Black Hole Information Loss Paradox}",
    volume = "118",
    year = "2017"
}

@article{Chen:2020sir,
    author = "Chen, Pisin and Mourou, Gerard",
    title = "{Trajectory of a flying plasma mirror traversing a target with density gradient}",
    eprint = "2004.10615",
    archivePrefix = "arXiv",
    primaryClass = "physics.plasm-ph",
    doi = "10.1063/5.0012374",
    journal = "Phys. Plasmas",
    volume = "27",
    number = "12",
    pages = "123106",
    year = "2020"
}

@article{unruh76,
  title = {Notes on black-hole evaporation},
  author = {Unruh, W. G.},
  journal = {Phys. Rev. D},
  volume = {14},
  issue = {4},
  pages = {870--892},
  numpages = {0},
  year = {1976},
  month = {Aug},
  publisher = {American Physical Society},
  doi = {10.1103/PhysRevD.14.870},
  url = {https://link.aps.org/doi/10.1103/PhysRevD.14.870}
}

@article{Bianchi:2014qua,
    author = "Bianchi, Eugenio and Smerlak, Matteo",
    title = "{Entanglement entropy and negative energy in two dimensions}",
    eprint = "1404.0602",
    archivePrefix = "arXiv",
    primaryClass = "gr-qc",
    doi = "10.1103/PhysRevD.90.041904",
    journal = "Phys. Rev. D",
    volume = "90",
    number = "4",
    pages = "041904",
    year = "2014"
}

@article{Larmor1897,
author = { J.   Larmor   D.Sc.   F.R.S. },
title = {LXIII. On the theory of the magnetic influence on spectra; and on the radiation from moving ions},
journal = {The London, Edinburgh, and Dublin Philosophical Magazine and Journal of Science},
volume = {44},
number = {271},
pages = {503-512},
year  = {1897},
publisher = {Taylor & Francis},
doi = {10.1080/14786449708621095}

}

@article{Barcelo:2010pj,
    author = "Barcelo, Carlos and Liberati, Stefano and Sonego, Sebastiano and Visser, Matt",
    title = "{Minimal conditions for the existence of a Hawking-like flux}",
    eprint = "1011.5593",
    archivePrefix = "arXiv",
    primaryClass = "gr-qc",
    doi = "10.1103/PhysRevD.83.041501",
    journal = "Phys. Rev. D",
    volume = "83",
    pages = "041501",
    year = "2011"
}

@inproceedings{Chen:2025xkv,
    author = "Chen, Pisin and Liu, Yung-Kun",
    title = "{Plasma wakefield: from accelerators to black holes}",
    eprint = "2509.03880",
    archivePrefix = "arXiv",
    primaryClass = "physics.plasm-ph",
    month = "9",
    year = "2025"
}

@article{Hsiung:2025mya,
    author = "Hsiung, Yang-Shuo and Chen, Pisin",
    title = "{A New Approach to the Calculation of Particle Creation from Analog Black Holes}",
    eprint = "2511.22895",
    archivePrefix = "arXiv",
    primaryClass = "gr-qc",
    month = "11",
    journal = {},
    year = "2025"
}

@article{Navick:2024wbd,
    author = "Navick, Xavier-Fran{\c{c}}ois",
    collaboration = "AnaBHEL",
    title = "{Design of the Setup for the AnaBHEL Experiment}",
    doi = "10.1007/s10909-023-03032-7",
    journal = "J. Low Temp. Phys.",
    volume = "214",
    number = "3-4",
    pages = "158--163",
    year = "2024"
}

@article{Dodonov:2025rxz,
    author = "Dodonov, Viktor V.",
    title = "{Dynamical Casimir Effect: 55 Years Later}",
    doi = "10.3390/physics7020010",
    journal = "MDPI Physics",
    volume = "7",
    number = "2",
    pages = "10",
    year = "2025"
}

@article{Good:2016oey,
    author = "Good, Michael R. R. and Anderson, Paul R. and Evans, Charles R.",
    title = "{Mirror reflections of a black hole}",
    eprint = "1605.06635",
    archivePrefix = "arXiv",
    primaryClass = "gr-qc",
    doi = "10.1103/PhysRevD.94.065010",
    journal = "Phys. Rev. D",
    volume = "94",
    number = "6",
    pages = "065010",
    year = "2016"
}

@article{Bianchi:2014vea,
    author = "Bianchi, Eugenio and Smerlak, Matteo",
    title = "{Last gasp of a black hole: unitary evaporation implies non-monotonic mass loss}",
    eprint = "1405.5235",
    archivePrefix = "arXiv",
    primaryClass = "gr-qc",
    doi = "10.1007/s10714-014-1809-9",
    journal = "Gen. Rel. Grav.",
    volume = "46",
    number = "10",
    pages = "1809",
    year = "2014"
}

@article{PisinChen2017 ,
  title = {Entropy evolution of moving mirrors and the information loss problem},
  author = {Chen, Pisin and Yeom, Dong-han},
  journal = {Phys. Rev. D},
  volume = {96},
  issue = {2},
  pages = {025016},
  numpages = {9},
  year = {2017},
  month = {Jul},
  publisher = {American Physical Society},
  doi = {10.1103/PhysRevD.96.025016},
  url = {https://link.aps.org/doi/10.1103/PhysRevD.96.025016}
}

@article{Davies:1974th,
    author = "Davies, P. C. W.",
    title = "{Scalar particle production in Schwarzschild and Rindler metrics}",
    doi = "10.1088/0305-4470/8/4/022",
    journal = "J. Phys. A",
    volume = "8",
    pages = "609--616",
    year = "1975"
}

@article{Fulling:1972md,
    author = "Fulling, Stephen A.",
    title = "{Nonuniqueness of canonical field quantization in Riemannian space-time}",
    doi = "10.1103/PhysRevD.7.2850",
    journal = "Phys. Rev. D",
    volume = "7",
    pages = "2850--2862",
    year = "1973"
}

@article{carlitz1987reflections,
  title = {Reflections on moving mirrors},
  author = {Carlitz, Robert D. and Willey, Raymond S.},
  journal = {Phys. Rev. D},
  volume = {36},
  issue = {8},
  pages = {2327--2335},
  numpages = {0},
  year = {1987},
  month = {Oct},
  publisher = {American Physical Society},
  doi = {10.1103/PhysRevD.36.2327},
  url = {https://link.aps.org/doi/10.1103/PhysRevD.36.2327}
}

@article{Lin:2021bpe,
    author = "Lin, Kuan-Nan and Chen, Pisin",
    title = "{Particle production by a relativistic semitransparent mirror of finite size and thickness}",
    eprint = "2107.09033",
    archivePrefix = "arXiv",
    primaryClass = "gr-qc",
    doi = "10.1140/epjc/s10052-024-12409-1",
    journal = "Eur. Phys. J. C",
    volume = "84",
    number = "1",
    pages = "53",
    year = "2024"
}

@article{Good:2022xin,
AUTHOR = {Ievlev, Evgenii and Good, Michael R. R.},
TITLE = "{Larmor Temperature, Casimir Dynamics, and Planck's Law}",
JOURNAL = {Physics},
VOLUME = {5},
YEAR = {2023},
NUMBER = {3},
PAGES = {797--813},
URL = {https://www.mdpi.com/2624-8174/5/3/50},
ISSN = {2624-8174},
eprint = "2211.00946",
    archivePrefix = "arXiv",
    primaryClass = "gr-qc",
DOI = {10.3390/physics5030050}
}

@article{AnaBHEL:2022sri,
    author = "Chen, Pisin and others",
    collaboration = "AnaBHEL",
    title = "{AnaBHEL (Analog Black Hole Evaporation via Lasers) Experiment: Concept, Design, and Status}",
    eprint = "2205.12195",
    archivePrefix = "arXiv",
    primaryClass = "gr-qc",
    doi = "10.3390/photonics9121003",
    journal = "Photon.",
    volume = "9",
    number = "12",
    pages = "1003",
    year = "2022"
}

@article{good2013time,
  title = {Time dependence of particle creation from accelerating mirrors},
  author = {Good, Michael R. R. and Anderson, Paul R. and Evans, Charles R.},
  journal = {Phys. Rev. D},
  volume = {88},
  issue = {2},
  pages = {025023},
  numpages = {16},
  year = {2013},
  month = {Jul},
  publisher = {American Physical Society},
  doi = {10.1103/PhysRevD.88.025023},
  url = {https://link.aps.org/doi/10.1103/PhysRevD.88.025023},  
     eprint={1303.6756},
   archivePrefix={arXiv},
   primaryClass={gr-qc}
}

@book{Jackson:490457,
      author        = "Jackson, John David",
      title         = "{Classical Electrodynamics}",
edition = "3rd ed.",
      publisher     = "Wiley",
      address       = "New York, NY",
      year          = "1999",
      url           = "https://cds.cern.ch/record/490457",
}

@article{Hawking:1974sw,
    author = "Hawking, S.W.",
    editor = "Gibbons, G.W. and Hawking, S.W.",
    title = "{Particle creation by black holes}",
    doi = "10.1007/BF02345020",
    journal = "Commun. Math. Phys.",
    volume = "43",
    pages = "199--220",
    year = "1975",
    }

@article{akal2022zoo,
    author = "Akal, Ibrahim and Kawamoto, Taishi and Ruan, Shan-Ming and Takayanagi, Tadashi and Wei, Zixia",
    title = "{Zoo of holographic moving mirrors}",
    eprint = "2205.02663",
    archivePrefix = "arXiv",
    primaryClass = "hep-th",
    reportNumber = "YITP-22-42, IPMU22-0023, YITP-22-42; IPMU22-0023",
    doi = "10.1007/JHEP08(2022)296",
    journal = "JHEP",
    volume = "08",
    pages = "296",
    year = "2022"
}

@article{ievlev2024moving,
    author = "Ievlev, Evgenii",
    title = "{Moving mirrors and event horizons in non-flat background geometry}",
    eprint = "2311.07403",
    archivePrefix = "arXiv",
    primaryClass = "gr-qc",
    reportNumber = "FTPI-MINN-23-21",
    doi = "10.1088/1361-6382/ad5bb4",
    journal = "Class. Quant. Grav.",
    volume = "41",
    number = "15",
    pages = "155009",
    year = "2024"
}

@article{Biswas:2024mlq,
    author = "Biswas, Parthajit and Ezhuthachan, Bobby and Kundu, Arnab and Roy, Baishali",
    title = "{Moving mirrors, OTOCs and scrambling}",
    eprint = "2406.05772",
    archivePrefix = "arXiv",
    primaryClass = "hep-th",
    doi = "10.1007/JHEP10(2024)146",
    journal = "JHEP",
    volume = "10",
    pages = "146",
    year = "2024"
}

@article{Hsiang:2024xlh,
    author = "Hsiang, Jen-Tsung and Hu, Bei-Lok",
    title = "{Foundational Issues in Dynamical Casimir Effect and Analogue Features in Cosmological Particle Creation}",
    eprint = "2410.03179",
    archivePrefix = "arXiv",
    primaryClass = "hep-th",
    doi = "10.3390/universe10110418",
    journal = "Universe",
    volume = "10",
    number = "11",
    pages = "418",
    year = "2024"
}

@article{Xie:2023wvu,
    author = "Xie, Yu-Cun and Butera, Salvatore and Hu, Bei-Lok",
    title = "{Optomechanical Backreaction of Quantum Field Processes in Dynamical Casimir Effect}",
    eprint = "2308.03129",
    archivePrefix = "arXiv",
    primaryClass = "quant-ph",
    doi = "10.5802/crphys.186",
    journal = "Comptes Rendus Physique",
    volume = "25",
    number = "S2",
    pages = "1--22",
    year = "2024"
}

@article{Gibbons:1976ue,
    author = "Gibbons, G. W. and Hawking, S. W.",
    title = "{Action Integrals and Partition Functions in Quantum Gravity}",
    doi = "10.1103/PhysRevD.15.2752",
    journal = "Phys. Rev. D",
    volume = "15",
    pages = "2752--2756",
    year = "1977"
}

@article{Hartle:1976tp,
    author = "Hartle, J. B. and Hawking, S. W.",
    title = "{Path Integral Derivation of Black Hole Radiance}",
    doi = "10.1103/PhysRevD.13.2188",
    journal = "Phys. Rev. D",
    volume = "13",
    pages = "2188--2203",
    year = "1976"
}

@article{Einstein:1918btx,
    author = "Einstein, Albert",
    title = {{{\"U}ber Gravitationswellen}},
    journal = "Sitzungsber. Preuss. Akad. Wiss. Berlin (Math. Phys. )",
    volume = "1918",
    pages = "154--167",
    year = "1918"
}

@article{Barcelo:2010xk,
    author = "Barcelo, Carlos and Liberati, Stefano and Sonego, Sebastiano and Visser, Matt",
    title = "{Hawking-like radiation from evolving black holes and compact horizonless objects}",
    eprint = "1011.5911",
    archivePrefix = "arXiv",
    primaryClass = "gr-qc",
    doi = "10.1007/JHEP02(2011)003",
    journal = "JHEP",
    volume = "02",
    pages = "003",
    year = "2011"
}

@article{Tomonaga:2024eoc,
    author = "Tomonaga, Masanori and Nambu, Yasusada",
    title = "{Second-order coherence as an indicator of quantum entanglement of Hawking radiation in moving-mirror models}",
    eprint = "2407.09218",
    archivePrefix = "arXiv",
    primaryClass = "quant-ph",
    doi = "10.1103/PhysRevD.110.105004",
    journal = "Phys. Rev. D",
    volume = "110",
    number = "10",
    pages = "105004",
    year = "2024"
}

@article{Planck1901,
  author    = {Max Planck},
  title     = {Ueber das Gesetz der Energieverteilung im Normalspectrum},
  journal   = {Annalen der Physik},
  volume    = {309},
  number    = {3},
  pages     = {553--563},
  year      = {1901},
  doi       = {10.1002/andp.19013090310}
}

\end{document}